\documentclass[12pt,thmsa,a4paper]{article}
\usepackage{sw20aip}

\input tcilatex
\QQQ{Language}{
American English
}

\begin{document}

\author{Hans J. Wospakrik\thanks{%
email: hansjw@fi.itb.ac.id} and Freddy P. Zen\thanks{%
Address after August 1, 1999: Optical Sciences Centre, Institute of Advanced
Studies, The Australian National University, Canberra, Australia.} \\
\\
Theoretical Physics Laboratory, \\
Department of Physics, Institute of Technology Bandung,\\
Jalan Ganesha 10, Bandung 40132, Indonesia}
\title{\textbf{CPT Symmetries and the B\"{a}cklund Transformations}}
\date{}
\maketitle

\begin{abstract}
We show that the auto-B\"{a}cklund transformations of the sine-Gordon,
Korteweg-deVries, nonlinear Schr\"{o}dinger, and Ernst equations are related
to their respective CPT\textbf{\ }symmetries. This is shown by applying the
CPT symmetries of these equations to the Riccati equations of the
corresponding pseudopotential functions where the fields are allowed to
transform into new solutions while the pseudopotential functions and the
B\"{a}cklund parameter are held fixed.
\end{abstract}

\section{Introduction}

It has been well-known that the usual (auto-) B\"{a}cklund transformation
(BT), of an integrable equation increases or decreases an integral number of
solitons. Precisely, within the inverse scattering method (ISM), BT changes
an integral number of poles and (or) zeros of the scattering data$^1.$ This
suggests a suspection that the BT\textbf{\ }may be related to the discrete
symmetries of the corresponding integrable equation, i.e., the \textbf{''}CPT%
\textbf{''} symmetries. Note that, by CPT\textbf{\ }symmetries, we mean for
either complex-conjugation (C), parity (P), time-reversal (T), or their
appropriate products.

It is the purpose of this paper to pursue this suspicion through some
specific examples. In fact, by using the CPT symmetries of the following
four popular integrable equations as case studies: the sine-Gordon (sG),
Korteweg-deVries (KdV), nonlinear Schr\"{o}dinger (NLS), and Ernst
equations, we are able to rederive their corresponding BT.

The key equation for the derivation of these BTs, by using this CPT\textbf{\ 
}method, is the Riccati equation of the corresponding pseudopotential
function. The derivation of the BT from this Riccati equation had also been
considered by Konno and Wadati as transformations that leave the Riccati
equation \textit{invariant}$^2.$ In \textit{contrast }to Konno-Wadati
approach, the present method does not \textit{preserve} the invariancy of
the Riccati equation under the above mentioned CPT transformations. This is
due to the ''rule'' that under these CPT transformations only the fields
that are allowed to transform into new solutions while the pseudopotential
functions and the B\"{a}cklund parameter are held unchanged. In this case,
the method could be considered as related to Chen's method$^3$. However, the
relation with the CPT symmetries was not explicitly stressed in Ref. 3.

In the next section, we will apply this method to derive the BT of the sG
equation. The BTs of the KdV, NLS, and Ernst equation will be derived in
Secs. 3, 4, and 5, respectively.

\section{The Sine-Gordon Equation}

The sine-Gordon equation is given by

\begin{equation}
\phi _{,uv}=\sin \phi ,  \tag{2.1}
\end{equation}
where $u=x+t,v=x-t$, and a comma preceding a subscript denotes partial
derivative. Henceforth, the $x$ and $t$ variables will be referred to the
space and time coordinates, respectively. The Riccati equations for the
corresponding pseudopotential function $\Gamma (u,v)$, as one derives by
using the Wahlquist-Estabrook prolongation method$^{4,5},$ read, 
\begin{equation}
\Gamma _{,u}=(\frac 12\lambda )(\Gamma ^2-1)\sin \phi +\lambda \Gamma \cos
\phi ,  \tag{2.2$a$}
\end{equation}
\begin{equation}
\Gamma _{,v}=\left( \frac 1\lambda \right) \Gamma +\frac 12\left( 1+\Gamma
^2\right) \phi _{,v},  \tag{2.2$b$}
\end{equation}
where $\lambda $ is the B\"{a}cklund parameter.

Following Chen$^3$, this set of Riccati equations may be considered as a
transformation between $\phi $ and $\Gamma $. Thus, \textit{if the same }$%
\Gamma $\textit{\ and }$\lambda $\textit{\ are also related to another
solution }$\widetilde{\phi },$ then Eq. $(2.2)$ provides a relation between
two solution $\phi $ and $\widetilde{\phi }$ of Eq. $(2.1)$, that is a 
\textit{B\"{a}cklund transformation}.

It is our purpose to find the other set of Ricatti equations for the same $%
\Gamma ,$ with the same $\lambda $, which express a transformation between
this same $\left( \Gamma ,\lambda \right) $ pair and the other solution $%
\widetilde{\phi }$ as the new solution that is generated by the discrete
''CPT\textbf{''} symmetries of the sG equation $(2.1)$.

Under the\textbf{\ }PT transformation: 
\begin{equation}
u\rightarrow u^{\prime }=-u,  \tag{2.3$a$}
\end{equation}
\begin{equation}
v\rightarrow v^{\prime }=-v,  \tag{2.3$b$}
\end{equation}
the sG equation $(2.1)$ is left invariant if the field $\phi $ transforms as
an eigenstate with eigenvalues $\pm $1, i.e., 
\begin{equation}
\phi (u,v)\rightarrow \phi ^{\prime }(u^{\prime },v^{\prime })=\pm \phi
(u,v).  \tag{2.4}
\end{equation}

Let us \textit{assume }that there exist a nontrivial transformed field $%
\widetilde{\phi }$ $(u,v)\neq \phi (-u,-v)$, such that, 
\begin{equation}
\phi (u,v)\rightarrow \phi ^{\prime }(u^{\prime },v^{\prime })=\pm 
\widetilde{\phi }(u,v),  \tag{2.5}
\end{equation}
is another solution of the sG equation $(2.1).$ In this case, the sG
equation $(2.1)$ will also be considered by us, as invariant under the PT
transformations $(2.3)$ and $(2.5)$. Our claim is that the relation between $%
\phi (u,v)$ and $\widetilde{\phi }(u,v)$ is given by the corresponding%
\textbf{\ }BT,\textbf{\ }as we will show by using the following method.

Since the new solution $\widetilde{\phi }(u,v)$ satisfies the sG equation $%
(2.1),$ i.e., 
\begin{equation}
\widetilde{\phi }_{,uv}=\sin \widetilde{\phi },  \tag{2.6}
\end{equation}
so within the Wahlquist-Estabrook prolongation method$^4$, we may choose to
hold the pseudopotensial function $\Gamma $ and the B\"{a}cklund parameter $%
\lambda $ unchanged under the PT transformations $(2.3)$ and $(2.5),$ i.e., 
\begin{equation}
\lambda \rightarrow \lambda ^{\prime }=\lambda ,\qquad \Gamma
(u,v)\rightarrow \Gamma ^{\prime }(u^{\prime },v^{\prime })=\Gamma (u,v). 
\tag{2.7}
\end{equation}

Equations $(2.5)$ and $(2.7)$ are essential to derive the corresponding
Riccati equations that connect the same $(\Gamma ,\lambda )$ pair of Eq. $%
(2.2)$ with the new solution $\widetilde{\phi }.$

However, it must be\ emphasized that since the CPT transformations are 
\textit{discrete}, i.e., they yield identity transformation when operated
twice, so the condition $(2.7)$ is only valid for two solutions, $\phi $ and 
$\widetilde{\phi }$, which are related by \textit{one particular} pair of $%
(\lambda ,\Gamma ).$

Thus, under the\textbf{\ }PT transformations $(2.3),$ $(2.5),$ and $(2.7)$,
the Riccati equations $(2.3a,b)$ become 
\begin{equation}
-\Gamma _{,u}=(\frac 12\lambda )(\Gamma ^2-1)\sin \widetilde{\phi }+\lambda
\Gamma \cos \widetilde{\phi },  \tag{2.8$a$}
\end{equation}
\begin{equation}
-\Gamma _{,v}=(\frac 1\lambda )\Gamma -\frac 12(1+\Gamma ^2)\widetilde{\phi }%
_{,v}.  \tag{2.8$b$}
\end{equation}

These Riccati equations could also be derived, using the Wahlquist-Estabrook
prolongation method, directly from Eq. $(2.6),$ in terms of the primed
coordinates $(u^{\prime },v^{\prime })$ and the field $\phi ^{\prime
}(u^{\prime },v^{\prime })$, with one condition that the PT transformation $%
(2.3)$ and the eigenvalue equation $(2.5)$ should only be applied to the
resulting Riccati equations.

Add Eq. $(2.2a)$ to $(2.8a)$ to eliminate $\Gamma _{,u},$ gives

\begin{equation}
\tan \frac 12(\widetilde{\phi }+\phi )=2\Gamma /(1-\Gamma ^2).  \tag{2.9}
\end{equation}
Solve for $\Gamma $, gives the following well-known relation between $\Gamma
,$ $\phi $ and $\widetilde{\phi },$%
\begin{equation}
\Gamma =\tan \frac 14(\widetilde{\phi }+\phi ),  \tag{2.10$a$}
\end{equation}
or 
\begin{equation}
\Gamma =-\text{cot}\frac 14(\widetilde{\phi }+\phi ).  \tag{2.10$b$}
\end{equation}

The Riccati equations $(2.2)$ together with Eq. $(2.10)$ constitute the BT
of the sG equation $(2.1)$. To reduce this BT\textbf{\ }into the usual form,
first of all we take the derivatives of Eq. $(2.10a),$ with respect to $u,$
and $v$, respectively. Use the Riccati equations $(2.2)$ again to eliminate $%
\Gamma _{,u}$ and $\Gamma _{,v}$ from the resulting equation, then after
performing simple trigonometric manipulations, we obtain the following
familiar looking BT\textbf{\ }of the sG equation$^4,$%
\begin{equation}
\widetilde{\phi }_{,u}=-\phi _{,u}+(2\lambda )\sin \frac 12(\widetilde{\phi }%
-\phi ),  \tag{2.11$a$}
\end{equation}
\begin{equation}
\widetilde{\phi }_{,v}=\phi _{,v}+(2/\lambda )\sin \frac 12(\widetilde{\phi }%
+\phi ).  \tag{2.11$b$}
\end{equation}

The relation (2.10\textit{b}) gives a similar transformation which differs
only in sign of the B\"{a}cklund parameter $\lambda .$

\section{The Korteweg-de Vries Equation}

The second equation we shall be considering in this section is the KdV
equation: 
\begin{equation}
\phi _{,t}+\phi _{,xxx}+12\phi \phi _{,x}=0.  \tag{3.1}
\end{equation}
The Riccati equations for the corresponding pseudopotential function $\Gamma
(x,t)$ are$^4$

\begin{equation}
\Gamma _{,x}=-(2\phi +\Gamma ^2-\lambda ),  \tag{3.2$a$}
\end{equation}

\begin{equation}
\Gamma _{,t}=4\left[ \left( \phi +\lambda \right) \left( 2\phi +\Gamma
^2-\lambda \right) +\frac 12\phi _{,xx}-\phi _{,x}\Gamma \right] , 
\tag{3.2$b$}
\end{equation}
where $\lambda $ is the B\"{a}cklund parameter.

The KdV equation (3.1) remains invariant under the PT transformation: 
\begin{equation}
x\rightarrow x^{\prime }=-x,  \tag{3.3$a$}
\end{equation}
\begin{equation}
t\rightarrow t^{\prime }=-t,  \tag{3.3$b$}
\end{equation}
\begin{equation}
\phi \rightarrow \phi ^{\prime }\left( x^{\prime },t^{\prime }\right) =%
\widetilde{\phi }\left( x,t\right) ,  \tag{3.3$c$}
\end{equation}
where $\widetilde{\phi }\left( x,t\right) $ is a new solution of the KdV
equation (3.1). Holding the pseudopotential function $\Gamma \left(
x,t\right) $ and the B\"{a}cklund parameter $\lambda $ unchanged under the
PT transformation (3.3), for the same reason as in Sec. 2, the Riccati
equations (3.2$a,b$) transform to 
\begin{equation}
-\Gamma _{,x}=-\left( 2\widetilde{\phi }+\Gamma ^2-\lambda \right) , 
\tag{3.4$a$}
\end{equation}
\begin{equation}
-\Gamma _{,t}=4\left[ \left( \widetilde{\phi }+\lambda \right) \left( 2%
\widetilde{\phi }+\Gamma ^2-\lambda \right) +\frac 12\widetilde{\phi }_{,xx}+%
\widetilde{\phi }_{,x}\Gamma \right] .  \tag{3.4$b$}
\end{equation}
Add Eq. (3.2$a$) to (3.4$a$) to eliminate $\Gamma _{,x}$ gives the well
known relation$^4$ : 
\begin{equation}
\widetilde{\phi }=-\phi -\Gamma ^2+\lambda .  \tag{3.5}
\end{equation}
Substraction, on the other hand, gives 
\begin{equation}
\Gamma _{,x}=\left( \widetilde{\phi }-\phi \right) ,  \tag{3.6}
\end{equation}
which is exactly Eq. (3.1-19) of Ref. 6, up to some constants.

Equations (3.2) and (3.5) constitute the BT of the KdV equation (3.1). In
fact, the usual form of this BT is obtained by eliminating the
pseudopotential function $\Gamma \left( x,t\right) $ from Eqs. (3.2) and
(3.5) then expressing the final result in terms of the potential $w\left(
x,t\right) $ through the relation$^4$, $\phi =-w,_x$.

\section{The Non-linear Schr\"{o}dinger Equation}

In this section, we will be considering the BT of the NLS equation, 
\begin{equation}
i\Psi _{,t}=-\Psi _{,xx}+\frac 12\varepsilon \Psi ^{*}\Psi ^2,  \tag{4.1$a$}
\end{equation}
where $\Psi $ is a complex field, and $^{*}$ is complex conjugation. Beside
the equation (4.1\textit{a}) we also have the complex conjugate equation: 
\begin{equation}
-i\Psi _{,t}^{*}=-\Psi _{,xx}^{*}+\frac 12\varepsilon \Psi \Psi ^{*2}. 
\tag{4.1$b$}
\end{equation}

The NLS equations (4.1$a,b$) remain invariant under the PT transformation, 
\begin{equation}
x\rightarrow x^{\prime }=-x,  \tag{4.2$a$}
\end{equation}
\begin{equation}
t\rightarrow t^{\prime }=-t,  \tag{4.2$b$}
\end{equation}
if the complex field $\Psi $ and its complex conjugate $\Psi ^{*}$ transform
according to the following rule: 
\begin{equation}
\Psi \left( x,t\right) \rightarrow \Psi ^{\prime }\left( x^{\prime
},t^{\prime }\right) =\widetilde{\Psi }^{*}\left( x,t\right) ,  \tag{4.3$a$}
\end{equation}
\begin{equation}
\Psi ^{*}\left( x,t\right) \rightarrow \Psi ^{*\prime }\left( x^{\prime
},t^{\prime }\right) =\widetilde{\Psi }\left( x,t\right) ,  \tag{4.3$b$}
\end{equation}
where $\widetilde{\Psi }^{*}\left( x,t\right) $ and $\widetilde{\Psi }\left(
x,t\right) $ are new solutions of Eqs. (4.1\textit{a}) and (4.1\textit{b}),
respectively. The transformation (4.3) is the anticipated \textit{C}
transformation. Thus, the NLS equation (4.1) remains invariant under the CPT
transformations (4.2) and (4.3).

The Riccati equations for the corresponding complex pseudopotential function 
$\Gamma \left( x,t\right) $ of Eq. (4.1), as derived by Estabrook and
Wahlquist, are$^6$%
\begin{equation}
2\Gamma _{,x}=-\Gamma ^2\Psi +\varepsilon \Psi ^{*}-2\lambda \Gamma 
\tag{4.4$a$}
\end{equation}
\begin{equation}
2\Gamma _{,t}=-i\lambda \left( \Gamma ^2\Psi -\varepsilon \Psi ^{*}-2\lambda
\Gamma \right) +i\left( -\Psi _{,x}\Gamma ^2-\varepsilon \Psi
_{,x}+\varepsilon \Psi ^{*}\Psi \Gamma \right)  \tag{4.4$b$}
\end{equation}
and for the complex conjugate pseudopotential function $\Gamma ^{*}\left(
x,t\right) $: 
\begin{equation}
2\Gamma _{,x}^{*}=-\Gamma ^{*2}\Psi ^{*}+\varepsilon \Psi -2\lambda
^{*}\Gamma ^{*},  \tag{4.4$c$}
\end{equation}
\begin{equation}
2\Gamma _{,t}^{*}=i\lambda ^{*}\left( \Gamma ^{*2}\Psi ^{*}-\varepsilon \Psi
-2\lambda ^{*}\Gamma ^{*}\right) -i\left( -\Psi _{,x}^{*}\Gamma
^{*2}-\varepsilon \Psi _{,x}^{*}+\varepsilon \Psi ^{*}\Psi \Gamma
^{*}\right) ,  \tag{4.4$d$}
\end{equation}
where $\lambda $ is the (complex) B\"{a}cklund parameter. Comparing with
Eqs. (9\textit{a}) and (9\textit{b}) for NLS in Ref. 2, one finds that, up
to the overall factor 2, their pseudopotential function $\Gamma $
corresponds to our complex conjugate pseudopotential function $\Gamma ^{*}$
in the coordinates $\left( -x,-t\right) $, and that their B\"{a}cklund
parameter is real. These differences are not essential since both lead to
the same BT of NLS, and we will use this freedom to fix our convention for
the transformation rules of a complex pseudopotential function $\Gamma $.

We observe that, since the new field $\Psi ^{\prime }\left( x^{\prime
},t^{\prime }\right) =\widetilde{\Psi }^{*}\left( x,t\right) $ satisfies the
NLS equation (4.1$b$), so within the Wahlquist-Estabrook prolongation method$%
^4$, we choose to interchange the pseudopotentials $\Gamma $ and $\Gamma
^{*} $ according to the rule: 
\begin{equation}
\Gamma \left( x,t\right) \rightarrow \Gamma ^{\prime }\left( x^{\prime
},t^{\prime }\right) =\Gamma ^{*}\left( x,t\right) ,  \tag{4.5$a$}
\end{equation}
\begin{equation}
\Gamma ^{*}\left( x,t\right) \rightarrow \Gamma ^{*\prime }\left( x^{\prime
},t^{\prime }\right) =\Gamma \left( x,t\right) ,  \tag{4.5$b$}
\end{equation}
as was used in Ref. 2.

In order to derive the usual relation between $\Gamma $ and $\Psi $ we fix
our convention by the following rule: ''If the complex field $\Psi $ and its
conjugate $\Psi ^{*}$ satisfy different equations, then under the CPT
transformation, the pseudopotential functions $\Gamma $ and $\Gamma ^{*}$
are chosen to obey the transformation rule (4.5), otherwise, they are kept
unchanged''.

Since $\Psi $ and $\Psi ^{*}$ satisfy different equations, i.e. (4.1$a$) and
(4.1$b$) respectively, so according to the above rule, we choose to
transform $\Gamma $ and $\Gamma ^{*}$ according to the rule (4.5).

Thus, under the CPT transformations (4.2), (4.3), and (4.5), the Riccati
equations (4.4$a$, $b$) transform to 
\begin{equation}
-2\Gamma _{,x}^{*}=-\Gamma ^{*2}\widetilde{\Psi }^{*}+\varepsilon \widetilde{%
\Psi }-2\lambda \Gamma ^{*},  \tag{4.6$a$}
\end{equation}
\begin{equation}
-2\Gamma _{,t}^{*}=-i\lambda \left( \Gamma ^{*2}\widetilde{\Psi }%
^{*}-\varepsilon \widetilde{\Psi }-2\lambda ^{*}\Gamma ^{*}\right) +i\left( 
\widetilde{\Psi }_{,x}^{*}\Gamma ^{*2}+\varepsilon \widetilde{\Psi }%
_{,x}^{*}+\varepsilon \widetilde{\Psi }\widetilde{\Psi }^{*}\Gamma
^{*}\right) ,  \tag{4.6$b$}
\end{equation}
and the Riccati equations (4.4$c$, $d$) to 
\begin{equation}
-2\Gamma _{,x}=-\Gamma ^2\widetilde{\Psi }+\varepsilon \widetilde{\Psi }%
^{*}-2\lambda ^{*}\Gamma ,  \tag{4.6$c$}
\end{equation}
\begin{equation}
-2\Gamma _{,t}=i\lambda \left( \Gamma ^2\widetilde{\Psi }-\varepsilon 
\widetilde{\Psi }^{*}-2\lambda ^{*}\Gamma \right) -i\left( \widetilde{\Psi }%
_{,x}\Gamma ^2+\varepsilon \widetilde{\Psi }_{,x}+\varepsilon \widetilde{%
\Psi }\widetilde{\Psi }^{*}\Gamma \right) .  \tag{4.6$d$}
\end{equation}
Note that the B\"{a}cklund parameter $\lambda $ is held unchanged under
these CPT transformations.

To derive the BT of the NLS equation, we first eliminate $\Gamma _{,x}$ and $%
\Gamma _{,x}^{*}$ from Eqs. (4.4\textit{a}), (4.4\textit{c}), (4.6\textit{a}%
), and (4.6\textit{c}). The final result is the following well-known relation%
$^{2,4}$: 
\begin{equation}
\widetilde{\Psi }=-\Psi -2\left( \lambda +\lambda ^{*}\right) \Gamma
^{*}/\left( \Gamma ^{*}\Gamma -\varepsilon \right)  \tag{4.7}
\end{equation}

Eqs. (4.4) and (4.7) constitute the BT of the NLS equation. The usual form
of this BT can be derived by solving Eq. (4.7) for $\Gamma $ then
substituting the result into the Riccati equation (4.4) to eliminate $\Gamma
_{,x}$ and $\Gamma _{,t}$. (The minus sign in front of the field $\Psi $ in
the right hand side of Eq. (4.7) is irrelevant since ($-\Psi )$ is also a
solution of Eq. (4.1)).

\section{The Ernst Equation}

In this final section, we proceed to apply this CPT method to rederive the
auto-BT of the celebrated Ernst's equation of the stationary axysymmetric
gravitational field or the static self-dual Euclidean SU(2) gauge field$^7$,
i.e. the Neugebauer \textit{I}$_1$ BT.

The Ernst equation is given by 
\begin{equation}
\left( \func{Re}\mathcal{E}\right) \left( 2V\mathcal{E},_{12}+V,_2\mathcal{E}%
,_1+V_{^{,}1}\mathcal{E},_2\right) =2V\mathcal{E},_1\mathcal{E},_2, 
\tag{5.1}
\end{equation}
where $\mathcal{E}$ is the complex Ernst potential, $\func{Re}\mathcal{E}$
the real part of $\mathcal{E}$. Henceforth, a comma preceding a subindex
denotes partial derivative with respect to the corresponding characteristic
coordinates : 
\begin{equation}
x^1=\rho +iz,\text{\qquad }x^2=\rho -iz,  \tag{5.2}
\end{equation}
with $\left( \rho ,z\right) $ are the cylindrical coordinates. $V$ is an
arbitrary harmonic function, i.e., 
\begin{equation}
V,_{12}=0.  \tag{5.3}
\end{equation}
Note that the complex conjugate Ernst potential $\mathcal{E}^{*}$ does also
satisfy the Ernst equation (5.1).

In terms of the Neugebauer field variables $\left( M_i,N_i\right) $, $%
i=1,2,3 $, introduced according to$^8$%
\begin{equation}
M_1=\left( 2\func{Re}\mathcal{E}\right) ^{-1}\mathcal{E},_1,\quad \text{ }%
M_2=\left( 2\func{Re}\mathcal{E}\right) ^{-1}\mathcal{E}^{*},_1,\text{ \quad 
}M_3=V^{-1}V,_1,  \tag{5.4}
\end{equation}
\[
N_1=\left( 2\func{Re}\mathcal{E}\right) ^{-1}\mathcal{E}^{*},_2,\qquad
N_2=\left( 2\func{Re}\mathcal{E}\right) ^{-1}\mathcal{E},_2,\qquad
N_3=V^{-1}V,_2, 
\]
the Ernst equation (5.1) together with its complex conjugate equation, and
the harmonic equation (5.3) are equivalent to the first order equations$^8$: 
\begin{equation}
M_{i,2}=C_i^{kl}M_kN_1,\quad \text{ }N_{i,1}=C_i^{kl}N_kM_1,  \tag{5.5$a$}
\end{equation}
where the non-vanishing $C_i^{kl}$ are given by 
\begin{equation}
C_1^{1\text{ }1}=C_2^{2\text{ }2}=C_3^{3\text{ }3}=-C_1^{1\text{ }2}=-C_2^{2%
\text{ }1}=-1,  \tag{5.5$b$}
\end{equation}
\[
C_1^{3\text{ }2}=C_1^{1\text{ }3}=C_2^{3\text{ }1}=-C_2^{2\text{ }3}=-\frac
12. 
\]

It is obvious that the Ernst equation (5.1), or the equivalent first order
equations (5.5) are invariant under the following CP symmetries:

(I) : 
\begin{eqnarray}
x^1 &\rightarrow &x^{1\prime }=x^2,\qquad \text{ }x^2\rightarrow x^{2\prime
}=x^1,  \tag{5.6} \\
\mathcal{E} &\rightarrow &\mathcal{E}^{\prime }\left( x^{1\prime
},x^{2\prime }\right) =\widetilde{\mathcal{E}}^{*}\left( x^1,x^2\right) , 
\nonumber \\
V &\rightarrow &V^{\prime }\left( x^{1\prime },x^{2\prime }\right) =%
\widetilde{V}\left( x^1,x^2\right) ,  \nonumber
\end{eqnarray}

(II) : 
\begin{eqnarray}
x^1 &\rightarrow &x^{1\prime }=x^2,\qquad x^2\rightarrow x^{2\prime }=x^1, 
\tag{5.7} \\
\mathcal{E} &\rightarrow &\mathcal{E}^{\prime }\left( x^{1\prime
},x^{2\prime }\right) =\widetilde{\mathcal{E}}\left( x^1,x^2\right) , 
\nonumber \\
V &\rightarrow &V^{\prime }\left( x^{1\prime },x^{2\prime }\right) =%
\widetilde{V}\left( x^1,x^2\right) ,  \nonumber
\end{eqnarray}
where $\widetilde{\mathcal{E}}$, and $\widetilde{\mathcal{E}}^{*}$ are new
solutions of Eqs. (5.1) or (5.5), and $\widetilde{V}$ of (5.3).

Apply the Wahlquist-Estabrook prolongation method to Eq. (5.5), one derives
the following Riccati equations$^{5,8}$ for the corresponding
pseudopotential functions $\alpha $ and $\gamma ,$ 
\begin{equation}
\alpha ,_1=\alpha \left( \alpha -1\right) M_1+\left( \alpha -\gamma \right)
M_2+\frac 12\alpha \left( \gamma -1\right) M_3,  \tag{5.8$a$}
\end{equation}
\begin{equation}
\alpha ,_2=\left( \alpha -1\right) N_1+\left( \frac \alpha \gamma \right)
\left( \alpha -\gamma \right) N_2+\left( \frac \alpha {2\gamma }\right)
\left( \gamma -1\right) N_3,  \tag{5.8$b$}
\end{equation}
\begin{equation}
\gamma ,_1=\gamma \left( \gamma -1\right) M_3,  \tag{5.8$c$}
\end{equation}
\begin{equation}
\gamma ,_2=\left( \gamma -1\right) N_3,  \tag{5.8$d$}
\end{equation}
For real solutions, $\alpha $, $\gamma $, and their complex conjugates
satisfy the relation, 
\begin{equation}
\alpha ^{*}=\alpha ^{-1},\text{ }\gamma ^{*}=\gamma ^{-1}.  \tag{5.9}
\end{equation}

Let us consider explicitly the CP transformation (5.6). Under these discrete
transformations, the Neugebauer field variables $M_i,$ $N_i$ transform
accordingly as follows: 
\begin{eqnarray}
M_1 &\rightarrow &\widetilde{N}_1,\qquad M_2\rightarrow \widetilde{N}%
_2,\qquad M_3\rightarrow \widetilde{N}_3,  \tag{5.10} \\
N_1 &\rightarrow &\widetilde{M}_1,\text{\qquad }N_2\rightarrow \widetilde{M}%
_2,\qquad N_3\rightarrow \widetilde{M}_3.  \nonumber
\end{eqnarray}

Since the Ernst fields $\mathcal{E}$ and its conyugate $\mathcal{E}^{*}$
satisfy the same field equation (5.1), so following our rule in Sec. 4, we
choose to hold $\alpha $ and $\gamma $ unchanged under this transformation.
Then, the Riccati equations $(5.8)$ transform to 
\begin{equation}
\alpha ,_2=\alpha \left( \alpha -1\right) \widetilde{N}_1+\left( \alpha
-\gamma \right) \widetilde{N}_2+\frac 12\alpha \left( \gamma -1\right) 
\widetilde{N}_3,  \tag{5.11$a$}
\end{equation}
\begin{equation}
\alpha ,_1=\left( \alpha -1\right) \widetilde{M}_1+\left( \frac \alpha
\gamma \right) \left( \alpha -\gamma \right) \widetilde{M}_2+\left( \frac
\alpha {2\gamma }\right) \left( \gamma -1\right) \widetilde{M}_3, 
\tag{5.11$b$}
\end{equation}
\begin{equation}
\gamma ,_2=\gamma \left( \gamma -1\right) \widetilde{N}_3,  \tag{5.11$c$}
\end{equation}
\begin{equation}
\gamma ,_1=\left( \gamma -1\right) \widetilde{M}_3.  \tag{5.11$d$}
\end{equation}

Eliminating $\alpha ,_1$, $\alpha ,_2$, $\gamma ,_1$, and $\gamma ,_2$ from
Eqs. (5.10) and (5.11), we derive: 
\begin{equation}
\left( \alpha -1\right) \left( \widetilde{M}_1-\alpha M_1\right) +\left(
\alpha -\gamma \right) \left( \frac \alpha \gamma \widetilde{M}_2-M_2\right)
=0,  \tag{5.12$a$}
\end{equation}
\begin{equation}
\left( \alpha -1\right) \left( \alpha \widetilde{N}_1-N_1\right) +\left(
\alpha -\gamma \right) \left( \widetilde{N}_2-\frac \alpha \gamma N_2\right)
=0,  \tag{5.12$b$}
\end{equation}
\begin{equation}
\widetilde{M}_3=\gamma M_3,\qquad \widetilde{N}_3=\left( \frac 1\gamma
\right) N_3.  \tag{5.13$a$}
\end{equation}
If the solutions are chosen to include the special cases $\alpha =1$, and $%
\alpha =\gamma $, then the (simplest) required solutions are: 
\begin{equation}
\widetilde{M}_1=\alpha M_1,\qquad \widetilde{M}_2=\left( \frac \gamma \alpha
\right) M_2,  \tag{5.13$b$}
\end{equation}
\[
\widetilde{N}_1=\left( \frac 1\alpha \right) N_1,\text{\qquad }\widetilde{N}%
_2=\left( \frac \alpha \gamma \right) N_2. 
\]
The transformation (5.13) is nothing but the Neugebauer $I_1$ BT$^8$.

The CP symmetry (5.7) does not lead to an essentially new BT since it is
nothing but (5.6) followed by complex conjugation on the new Ernst
potentials $\widetilde{\mathcal{E}}$ and $\widetilde{\mathcal{E}}^{*}$ or 
\textit{vice versa}.

\section*{Concluding Remarks}

In conclusion, we have shown that the auto-B\"{a}cklund transformations of
the sG, KdV, NLS, and Ernst equations are related to their respective CPT
symmetries. In fact, the above analysis shows that the explicit field
equation is prerequisite for the application of this CPT method for deriving
the auto-BT of a two dimensional integrable equation.

\section*{Acknowledgements}

It is a pleasure for us to acknowledge the Directorate General of Higher
Education of the Republic of Indonesia, for supporting this research under
the project: Hibah Bersaing VII/2, 1999-2000. We also thank Drs. M. Husin
Alatas and Bobby E. Gunara for discussions, Dr. Alexander P. Iskandar for
the comment, and Prof. P. Silaban for the encouragement. One of us, (F. P.
Z) would like to thank YANBINBANG-SDM IPTEK (Habibie Foundation) for
financial support.

\end{document}